# Pot, kettle: Nonliteral titles aren't (natural) science[1]

Mike Thelwall, Statistical Cybermetrics Research Group, University of Wolverhampton, UK.

Researchers may be tempted to attract attention through poetic titles for their publications, but would this be mistaken in some fields? Whilst poetic titles are known to be common in medicine, it is not clear whether the practice is widespread elsewhere. This article investigates the prevalence of poetic expressions in journal article titles 1996-2019 in 3.3 million articles from all 27 Scopus broad fields. Expressions were identified by manually checking all phrases with at least 5 words that occurred at least 25 times, finding 149 stock phrases, idioms, sayings, literary allusions, film names and song titles or lyrics. The expressions found are most common in the social sciences and the humanities. They are also relatively common in medicine, but almost absent from engineering and the natural and formal sciences. The differences may reflect the less hierarchical and more varied nature of the social sciences and humanities, where interesting titles may attract an audience. In engineering, natural science and formal science fields, authors should take extra care with poetic expressions, in case their choice is judged inappropriate. This includes interdisciplinary research overlapping these areas. Conversely, reviewers of interdisciplinary research involving the social sciences should be more tolerant of poetic license.
**Keywords:** Journal article titles; Disciplinary differences; Academic humor. Poetic titles.

## 1 Introduction

Journal article titles are a vital component of scholarship. They are likely to be the first thing read during literature searches and journal browsing, potentially triggering an initial decision to read or ignore the associated article. In addition, the words in the title form part of the indexing of the article, affecting how it can be found through digital library searches. Titles may also give an initial impression to reviewers, influencing their overall judgement. Because of these factors, constructing an appropriate title is an important scholarly skill. In the context of the increasing amount of interdisciplinary research and scholars that are not native English speakers (needing training: Kuteeva & Negretti, 2016), it is useful to investigate different aspects of how article titles may be constructed.

This article focuses on common poetic expressions in article titles. Poetic expressions can attract attention but may create a negative impression if inappropriate for a field, may lose a search audience if not keyword-rich, and clichés may create a negative impression on readers (Lindauer, 1968).

Despite the mainly biomedical investigations of standard title phrases (reviewed below), there have been no science-wide studies of poetic or clichéd article titles yet, and all previous studies have started with lists of candidate phrases (e.g., Shakespeare plays) rather than seeking evidence about which phrases are common. These are important omissions because, there is no theory-driven reason yet to believe that one type of poetic expression would be more common, and there are disciplinary differences in the way knowledge is constructed and written about (Hyland, 2012; Whitley, 2000). In particular, the natural sciences are more hierarchical (Kuteeva & Airey, 2014) and the social sciences refer more to previous publications (Hyland, 1999). This article addresses the interdisciplinary and list-

---

based research gaps with a science-wide study of common stock phrases extracted from science-wide article titles with a heuristic and manual checking rather than pre-defined lists.

## 2 Idioms and poetic language

Language can be analysed from the artistic perspective of poetry. This form of writing uses a variety of methods (poetic devices) to create meaning, rhythm or mood. The many recognised devices include idioms, alliteration, assonance and allusion, as well as metaphor and simile, all of which also occur in other types of text (Chovanec, 2008). A metaphor, allusion or poetic phrase is most powerful when first coined but weakens over time to become a cliché. Metaphors eventually fall out of use or "die" (Alm-Arvius, 2006; Lakoff, 1987) by becoming an accepted meaning (e.g., "I *see* what you mean" is no longer a metaphor for understanding: Geary, 2011).

Particularly relevant here, an allusion is a figure of speech that makes covert reference to another object, forcing the reader to decode the connection (Perri, 1978). This object alluded to could be a person or place, but it could also be another form of text, such as a popular song lyric. The allusion might be central to the meaning or add an extra dimension (Leppihalme, 1997). Allusions do not seem to have been systematically studied in a way that is relevant here, but other types of poetic device, and their use in academia, are discussed below.

### *2.1 Idioms*

Idioms have been studied from a linguistic perspective, such as to understand how language works or to support natural language processing and language teaching. Although the concept of an idiom is well known and previous investigations have distinguished between different types, there is not an agreed definition (Espinal & Mateu, 2010; Nunberg, Sag, & Wasow, 1994). In general, idioms are expressions that are commonly used as a single unit and have a figurative meaning. Whilst the meanings of many idioms are unrelated to the individual words (e.g., "Bob's your uncle" in British English suggests that success is assured), more conventional idioms have a literal meaning that is related to their figurative meaning (e.g., "close to the bone"). An expression can become idiomatic when it continues to be used after its original meaning has been forgotten (e.g., "Bob" above was probably a nepotistic 19[th] century British politician Robert Cecil). A recent study suggests that modifying idiomatic phrases that are usually fixed does not obscure their meaning (Kyriacou, Conklin, & Thompson, 2019), and this device may lengthen the life of an idiom when appropriately used.

Idioms are a type of multi-word expression (Huening & Shlucker, 2015), and a type of collocation: a sequence of words occurring more often than statistically expected (Clear, 1993). Non-idiomatic multi-word expressions may still have idiosyncratic elements, such as employing one or more archaic words or meanings (e.g., "spick and span", "stand and deliver").

An irreversible binomial expression is a common type of multi-word expression. It is a fixed order sequence of two words or groups of words with the same part of speech (e.g., both verbs or both adjectives), connected by "or" or "and". Irreversible binomials are often alliterative (e.g., "to have and to hold", "short and sweet") or rhyming (e.g., "high and dry"), thus injecting them with a poetic element that presumably contributes to their widespread use. Some, but not all, irreversible binomials are also idioms (e.g., "rich and famous" vs. "do or die").

## 2.2  Sayings, proverbs, aphorisms and clichés

Proverbs are "short sentences of wisdom" (Mieder, 2004, p.3) that are well known and have persisted in a cultural context, such as "don't judge a book by its cover". Due to their short nature, they are susceptible to incorporation within sentences or titles. In contrast, an aphorism is a memorable concise expression of a general truth, possibly with an element of metaphor, such as "all that glitters is not gold" (McGlone & Tofighbakhsh, 2000). More generally, a saying can be defined as any memorable concise expression. These are not poetic devices but are poetic language in the more general sense.

A recognisable expression can be refreshed or injected with humour by modifying it in various ways, including by shortening, substituting individual words or putting it in a novel context (Vrbinc & Vrbinc, 2011).

A cliché is a figurative expression that is perceived as being overused and that is likely to drop from common usage as a result (Lindauer, 1968). Clichés can generate a bad impression, including in academic writing, and this may extend to common stylistic expressions (Deb, Dey, & Balas, 2019).

## 2.3  Poetic language in academia

Previous studies have noted the prevalence of song lyrics in journal article titles, employing the poetic device of allusion. Some prior studies have also analysed poetic or idiomatic language in other academic contexts.

An analysis of idiomatic expressions in 152 transcribed spoken academic texts from a US university between 1997 and 2001 found that many were commonly used, and would be helpful for second language learners to understand. These included, "chicken-and-egg", "the big picture" and "in a nutshell" (Simpson & Mendis, 2003), which would presumably help to add interest to a lecture but might be too informal for routine use in journal articles.

One reasonably discipline-specific analysis of journal articles investigated the use of figurative language in articles about the cloning of the sheep Dolly (Giles, 2001). It found extensive use of many types, including metaphor, cliché and hyperbole, but not an agreed central metaphor for the issue.

Poetic language can become deeply embedded into some disciplines. In computer science, metaphors are essential to help convey complex abstract concepts (Colburn & Shute, 2008), such as "bag of words" for the unordered set of all words in a text (also used in computational linguistics). Metaphors in common use include Windows (function metaphor), menu (function metaphor), email (function metaphor) and mouse (appearance metaphor). In addition, computer scientists (both academic and non-academic) sometimes adopt apparently playful names, such as relating to animals. The many examples include Python (programming language; primarily alluding to Monty Python's Flying Circus[2]), GNU (recursive acronym for "GNU's Not Unix!"), bug (metaphor for programming error), Gopher (metaphor for search helper), Panther (Apple operating system codename), and ASP (Active Server Pages). Spam was originally a playful computer science allusion to a Monty Python's Flying Circus comedy sketch from 1970. It has become a serious term and literal description with its own academic conferences (Conference on Email and Anti-Spam: CEAS) and is mentioned in many paper titles (e.g., "Web spam detection using SVM classifier").

Other fields may also inject humour or interest through names when a new object is created or identified. Physics examples arguably include strangeness, charm, and flavor (of quark). In biology, animal names include the fried egg jellyfish, the hebejeebie plant genus and the aha ha wasp. Chemical compound names with allusions include the proteins ranasmurfin (Smurfs), pikachurin (Pikachu) and sonic hedgehog. Black swan theory in economics uses a literary allusion. In architecture, there are moon, pigtail and roving bridges. Information science allusions include sleeping beauty (now delayed recognition) and Mathew effect (also known as rich-get-richer). These cases all seem to serve the purpose of making the named object memorable or attracting attention to it or the associated documentation.

---

[2] https://docs.python.org/2/faq/general.html#why-is-it-called-python

## 2.4 Article titles

Journal article titles have been previously studied mainly for their abstract informational value (Diener, 1984). Statistical analyses of various fields have found a range of title-related factors to associate with increased citation impact. These factors include the title length (cf. Didegah & Thelwall, 2013), and specific acronyms, with negative factors including country names and questions, and factors that could be positive or negative (depending on the study) including the presence of a colon (Jacques & Sebire, 2010; Jamali, & Nikzad, 2011). Questions in titles may also be used to attract an audience (Ball, 2009; Cook & Plourde, 2016), as may titles that do not overlap with an article's keywords (Rostami, Mohammadpoorasl, & Hajizadeh, 2014). A study of many title properties for management science found two to have a small association with citation counts (Nair & Gibbert, 2016). It is difficult to justify a cause-and-effect relationship for these factors although it seems reasonable that articles with country names tend to be less cited because they have a more specific focus. There are disciplinary differences in the construction of titles, including their average length and the proportion of substantive (content-bearing) words (Nagano, 2015), as well as the use of colons (Hartley, 2007). There are also disciplinary differences and historical evolutions in the lengths and structure of titles (Milojević, 2017). In the social sciences, the nationality of authors may influence their titles (Kim, 2015). Quotes from survey or interview respondents may also be used in titles (Pułaczewska, 2009). These all may perform the similar function of attracting attention.

Poetic language is sometimes used in article titles. In the biomedical literature, the use of Shakespearean quotes (e.g., "To be or not to be"), film titles (e.g., "Back to the future"), and fairy stories (e.g., "The Emperor's New Clothes") in article titles increased between 1950-54 and 2000-04 (Goodman, 2005). Biomedical clichés subsequently found include: state of the art; gold standard; paradigm shift/s/ing; cutting edge; outside the box; wind/s of change; coalface; goalposts; and playing field (Goodman, 2012). Bob Dylan lyrics can also be found in the biomedical literature (e.g., "Like a rolling histone: epigenetic regulation of neural stem cells and brain development by factors controlling histone acetylation and methylation"), partly as the result of a bet in Sweden (Gornitzki, Larsson, & Fadeel, 2015). Bob Dylan songs are also exploited by meteorologists (Brown, Aplin, Jenkins, Mander, Walsh, & Williams, 2016). Numerous medical papers have taken advantage of the acronym of the National Institute for (Health and) Clinical Excellence to make clever titles (e.g., "The NICE cost-effectiveness threshold") (Morrison & Batty, 2009), and sex-related titles seem to gain extra readers (Langdon-Neuner, 2008).

There is mixed evidence on whether attempts at poetic titles tend to be successful. Medical papers with clichés in their title seem more likely to be rejected (Gjersvik, Gulbrandsen, Aasheim, & Nylenna, 2013). In psychology, articles with titles rated amusing by human judges tend to be more cited (Subotic, & Mukherjee, 2014). In contrast, a *Biological*

*Conservation* editorial argued that title characteristics have little impact on citation rates in the journal (Costello, Beard, Primack, Devictor, & Bates, 2019).

## 3 Methods

Common poetic expressions were sought by extracting word n-grams from all Scopus article titles 1996-2019. Before counting, all words were converted to lower case and punctuation was removed. Titles were split at colons or full stops to prevent the end of one phrase from being merged with the start of another. The following rules were used to select a candidate list of common expressions.

- The n-gram must contain at least 5 (consecutive) words. This is a practical step to produce a manageable list of phrases to check.
- The n-gram must occur in at least 25 titles. This is again a practical step to produce a manageable list of phrases to check.
- The n-gram must not be part of a longer n-gram that occurs at least 50% as often. This is a heuristic to focus on the longest form of an expression.

This procedure is available in the free software Webometric Analyst (lexiurl.wlv.ac.uk in the submenu: *Tab-sep|Count frequency of texts or words in column|Find longest Ngrams in text or column*, then selecting the title column).

I manually checked the 152,928 n-grams found to seek phrases that were poetic, excluding noun phrases and non-poetic common phrases (e.g., "The purpose of this article is to"). I subjectively judged a phrase as *poetic* if it *appeared to use words that were unnecessary to convey the meaning succinctly and clearly*. This included the use of extra words, unusual words (e.g., "tale" as an old-fashioned word for story, "kids" as an informal word for children, "you" as an unlikely personal pronoun), or idioms. It also included phrases that seemed unlikely to be widely useful in academic research at face value (e.g., "on the road to a").

The judgements were time consuming and error prone due to the length of the task and the possibility of poetic phrases from cultural contexts unknown to me. Even though I almost certainly missed some poetic phrases (in addition to shorter expressions that were excluded by design), the phrases found represent a large enough set to allow disciplinary comparisons. The full list of n-grams can be found online (https://figshare.com/articles/Ngrams_and_figures_for_Funny_titles_aren_t_natural_science/12016278). This subjective method was chosen in preference to the approach used in previous quantitative studies that matched titles to pre-defined candidate lists (e.g., Shakespeare quotes, Queen lyrics) to avoid pre-determining the types of phrases that might be found. Identifying modified idiomatic phrases automatically is challenging (Weber, Fischer & Dormeyer, 2007). An advantage of this approach is its ability to detect modified phrases, such as "to screen or not to screen". Nevertheless, it has the substantial alternative limitation of subjectivity. In particular, the sources of phrases might be misinterpreted due to a lack of knowledge of subject-specific canonical texts, or of texts not known to a person of my cultural background and experiences (e.g., most U.S. country music, songs originally not in English, traditional Chinese sayings, classical music). This is particularly likely for phrases that have a plausible surface meaning coupled with an allusion (e.g., "the curious case of the").

After the initial detection of apparently poetic phrases, the selected n-grams were re-examined to make a consistent decision about whether a phrase was poetic or just standardised. Some phrases were at the borderline and judgement was needed. These decisions were made before any analyses so that the results could not influence disciplinary differences. For example, "a race to the bottom" was kept but "a view from the trenches" was rejected. The latter was repeatedly used in the specific context of an article or column describing the practitioner perspective, making it seem (to me) to be a standardised title-specific expression rather than a less predictably used expression that readers might not expect to see in an article title.

The titles of articles containing the initially selected phrases were examined and the phrases were rejected when most articles were about associated works (e.g., "The picture of Dorian Grey", "The turn of the screw", "The lord of the rings", "Their eyes were watching god") or when the phrase did not seem to be poetic within its discipline (e.g., "across the tree of life", "doing well by doing good" [a business concept]). After the filtering, 149 common poetic phrases remained.

For additional context, the 149 phrases were classified into several types, according to their (guessed) most common use, as defined here. These include literary or cultural allusions, idioms and proverbial expressions. Approximate matches were allowed when the modifications were judged to be alluding to the unmodified form (e.g., "a tale of two cultures", presumably derived from, "A tale of two cities").

- Literary allusion: A book, poem or play title or quote (e.g., Shakespeare).
- Film: The name of a film. No film quotes were identified (e.g., "You're gonna need a bigger boat") but would have been included.
- Music: A musical title or song lyric.
- Saying: A well-known English proverbial phrase.
- Stock phrase: A common idiom or other phrase that is neither proverbial nor an allusion.

In many cases a phrase could be in multiple classes. For example, most of the phrases are probably in book titles, if relatively unknown volumes are included. Some phrases were apparently variations on well-known phrases and were classified according to the type of the associated phrase (e.g., "to eat or not to eat" as a variation of "to be or not to be"; Other phrases alluded to a book title, such as "A tale of three cities" as a variation on "A tale of two cities"). Others omitted a key word but kept the structure of an associated phrase (e.g., "the dark side of the [moon]", [the king/queen] is dead, long live the [king/queen]").

The relative prevalence of poetic phrases across disciplines was assessed by calculating the percentage of articles in each Scopus broad category (n=27) with the phrases in their titles. Articles classified in more than one broad category were counted fractionally. For example, an article in two broad categories would count as 0.5 articles for both. The same fractional counting procedure was used for the number of matching phrases and the number of articles in each category. Eight phrases were selected to report in detail for additional context.

## 4  Results

The 149 common phrases were contained in between 25 and 477 article titles. The article titles sometimes used the phrases in a context in which they were accurate descriptions, with the poetry deriving from the choice of language. For example, titles including the phrase "a tale of two cities" typically involved a comparison between two geographic locations (often cities), and this was a poetic way to allude to this. The phrase is a Charles Dickens novel, otherwise the phrase "a tale" would be old fashioned and inappropriate for a journal article not focusing on a story narrative.

The extent to which the 149 poetic phrases occur in the 27 Scopus broad fields varies greatly (Figure 1). Whilst more than 1 in 900 articles contain one of them in six broad fields, they occur in less than 1 in 29,000 Materials Science journal articles. At the extreme difference level, they are 44 times more likely to occur in a Social Sciences article title than in a Materials Science article title, although they are rare in both. The delineations follow the Scopus Social Sciences and Arts and Humanities categories, although there may well be substantial variation within them. The data behind the figures can be found online (https://figshare.com/articles/Ngrams_and_figures_for_Funny_titles_aren_t_natural_science/12016278).

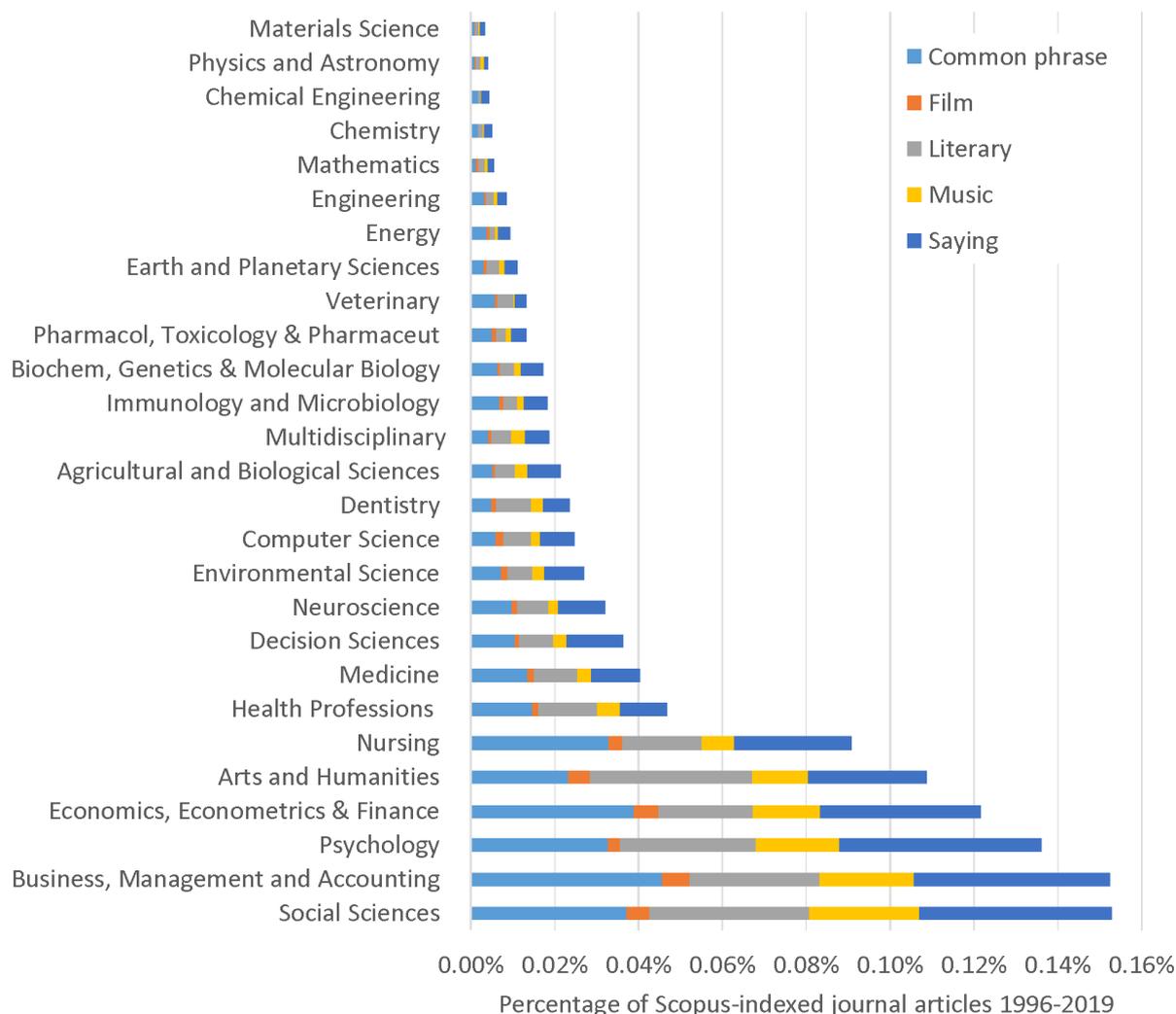

Fig 1. The percentage of Scopus-indexed journal articles 1996-2019 containing one of the 149 poetic phrases identified by Scopus broad category. All counting is fractional for articles in multiple broad fields.

Many different stock phrases were reasonably common (Figure 2). The origins are varied, from commercial ("does one size fit all", "two for the price of one") to biblical ("in search of the holy grail"), Christian weddings ("till death do us part"; also a UK TV show), and light metaphors ("a seat at the table"). Some may be influenced by television program names ("who wants to be a [millionaire]") or TV catchphrases ("2 for the price of 1" in The Price is Right), or less well known songs (Abba: "Two for the price of one").

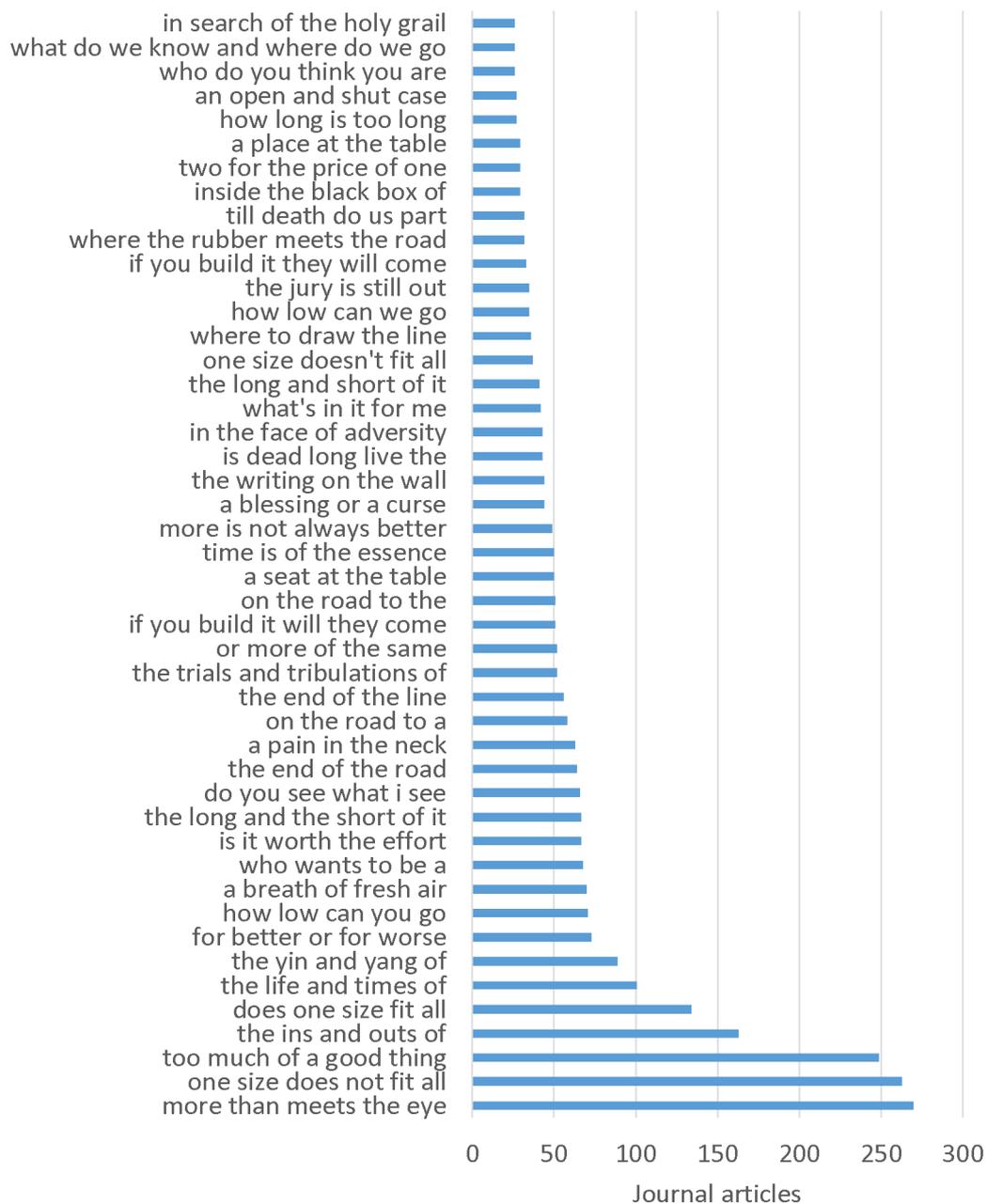

Fig 2. Stock phrases with at least 5 words in Scopus journal article titles 1996-2019.

The literary allusions (Figure 3) are full or partial book titles (including plays), famous quotes from books (or plays, usually Shakespeare), and word play variations of them. One of the books is a ground-breaking popular 1963 children's story (Where the wild things are) and one is over 2,500 years old (The tortoise and the hare). The only post-1996 book is "Catch me if you can", but this is perhaps more famous as a film, and the phrase was in common use before this (e.g., a 1969 film, songs, and a 165 play).

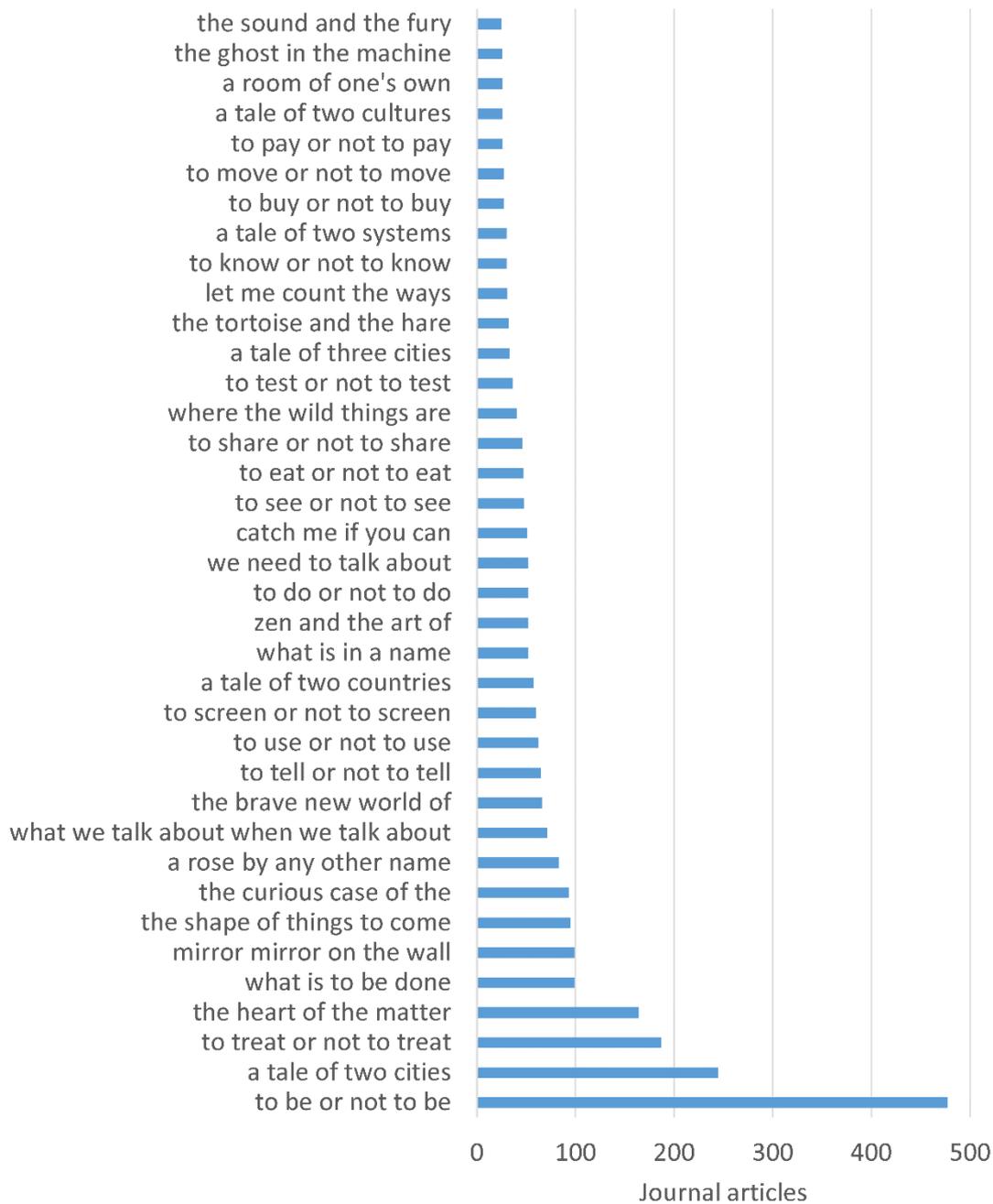

Fig 3. Literary allusions with at least 5 words in Scopus journal article titles 1996-2019.

There seemed to be only three film titles included (Figure 4), including "how I learned to stop worrying and love [the bomb]", which is associated with the film Dr Strangelove. The phrase "what are they good for" seems to be a play on the Motown song *War*, containing the chorus, "War, what is it good for? Absolutely nothin'!". The Tina Turner song "What's love got to do with it" is in twice, once for the apparently truncated part "got to do with it". This was commonly used in the title phrase, "What's [x] got to do with it," where [x] is the topic of the article. One short phrase seems to be an allusion to a Pink Floyd song, "The dark side of the [moon]", although this could be a coincidence since it could also be a plain description. Two phrases may be influenced by pop group New Kids on the Block (active 1985-1994), although this is a common expression.

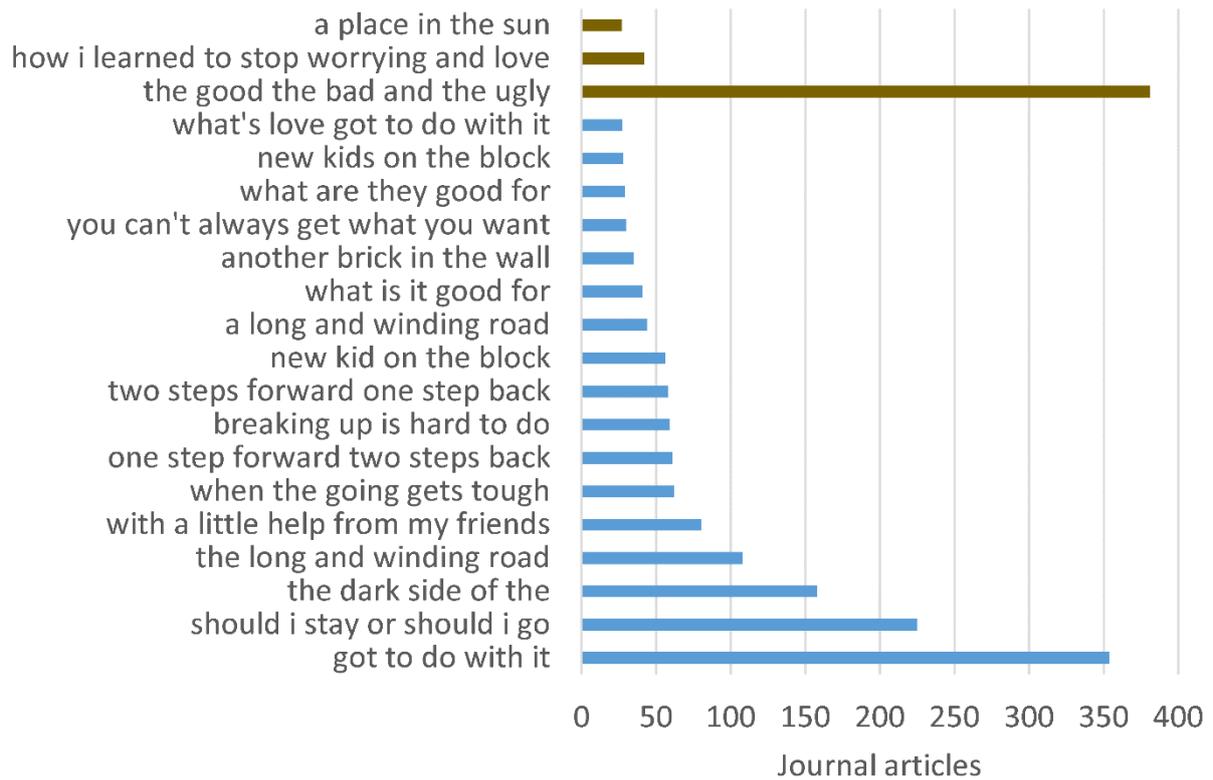

Fig 4. Song (bottom) or film (top) titles or allusions with at least 5 words in Scopus journal article titles 1996-2019.

Many common sayings are widely used, including with several variations of the same saying (Figure 5). Some are incomplete sayings, but allude to the original sentence (e.g., "more than one way to [skin a cat]", "[beauty] is in the eye of the beholder").

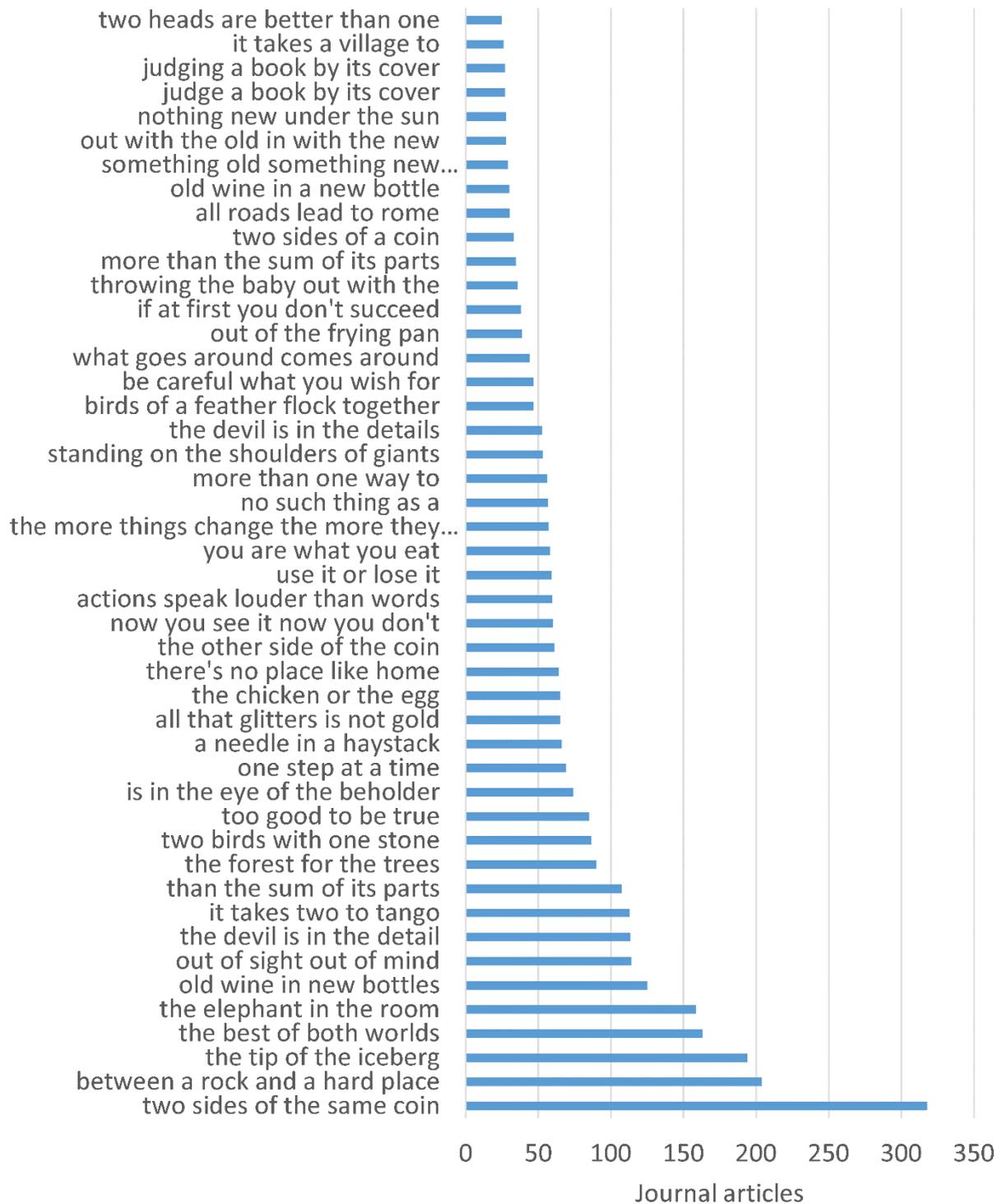

Fig 5. Sayings with at least 5 words in Scopus journal article titles 1996-2019.

Eight selected expressions were investigated further for background information on use contexts. These expressions were chosen non-randomly to illustrate a range of different types.

- **Two sides of the same coin**: This traditional saying was commonly used to express something with positive and negative aspects in the social sciences and medicine. In Education, an example title is, "Competences for learning to learn and active citizenship: Different currencies or two sides of the same coin?". In Medicine, the articles seemed to be overviews, short reports or case studies rather than primary research ("Tumor dormancy and cancer stem cells: two sides of the same coin?"). An example of a rare life sciences use is, "Overexpression of protein phosphatase 5 in the mouse heart: Reduced contractility but increased stress tolerance – Two sides of the same coin?".

- **Between a rock and a hard place**: This saying started 28 articles (not in a series) in the Sociology and Political Science narrow field (part of Social Science) ("Between a rock and a hard place: Radical Islam in post-Suharto Indonesia"). It is used in political contexts when there are problems with only difficult options. No titles used the phrase literally, with "Between a rock and a hard place: Environmental and engineering considerations when designing coastal defence structures" being the closest.
- **The good the bad and the ugly**: This Italian Spaghetti Western movie title suggests a study with three relevant factors, with one positive, one negative and one very negative. In Medicine, the phase seemed to be used in non-primary research general articles ("The good, the bad and the ugly of federal health care reform"). Articles sometimes explicitly referred to three things in the title ("The good, the bad and the ugly: Three faces of social media usage by local governments") or abstract ("The 'good', the 'bad' and the 'ugly'? views on male teachers in foundation phase education" referred to liked, disliked and threatening teachers). For other papers, the metaphor is not strongly enough tied to the content of the paper to be understandable from the abstract (e.g., the abstract of "Big Data for Policy Analysis: The Good, The Bad, and The Ugly" only mentions the existence of positive and negative factors.
- **The elephant in the room**: This saying apparently originated with a nineteenth century Russian fable. The matching titles of papers were implicit arguments that an aspect of the paper is well known but avoided as a topic of discussion or research ("Lateral violence: Calling out the elephant in the room", "Race and research in the southern United States: Approaching the elephant in the room", "The elephant in the room: Poverty, disability, and employment").
- **Old wine in new bottles**: This saying (with biblical origins) might suit the title of an article assessing whether a new development or essentially the same as before, and was useful for political contexts ("Old wine in new bottles? The 1999 Finnish election campaign on the internet", "Old Wine in New Bottles? What is the Tea Party and Where Did it Come From?"), including for social geography ("Human development - new paradigm or old wine in new bottles?").
- **Out of sight out of mind**: This traditional saying was used when a topic involved forgetting as a component, particularly in medicine ("Foot trauma due to foreign bodies - Out of sight, out of mind?"), or ignoring something ("Out of sight, out of mind: how Harvard University exploited rural Chinese villagers for their DNA."), but also in sociology and psychiatry ("Out of sight, out of mind: Workplace smoking bans and the relocation of smoking at work").
- **The long and winding road**: This Beatles song is used to emphasize the length of time needed for a project ("The long and winding road. Arriving at safe medication management in LTC setting", "The long and winding road to personalized glycemic control in the Intensive Care Unit") but sometimes to refer to twisting ("Epidural catheters: The long and winding road").

- **One step forward two steps back**: This Desert Rose Band song title is used to highlight that some progress has been made but the situation has got worse, particularly in a political context ("One step forward, two steps back: Success and failure in recent Turkish foreign policy", "One step forward, two steps back: The political culture of corruption and cleanups in Nigeria"). A reviewer of this article pointed out that this is the title of a book by Lenin, which is the probable motivation behind its use within most or all article titles in politics. This shows the importance of subject-specific knowledge in making interpretations and that some of my classifications are at least partly wrong. In this case, other uses seemed unlikely to be politics-motivated (e.g., "Pre-operative staging of breast cancer with breast MRI: One step forward, two steps back?", "One step forward, two steps back - Will there ever be an AIDS vaccine?", and "One step forward, two steps back: Arguing for a transatlantic investor protection regime"), so this phrase probably has multiple motivations.

# 5  Discussion

This paper is limited by the focus on common poetic phrases and analyses their prevalence rather than their effectiveness. Paper titles may also be poetic or idiomatic through modified quotes ("You probably think this paper's about you: narcissists' perceptions of their personality and reputation", "miR miR on the wall, who's the most malignant medulloblastoma miR of them all?",  "Fantastic yeasts and where to find them: the hidden diversity of dimorphic fungal pathogens"), rarer phrases ("Love will tear us apart: transformational leadership and love in a call centre"), cursing ("Fk yea I swear: cursing and gender in MySpace"), greatly modified poetic phrases ("Just what is critical race theory and what's it doing in a nice field like education?"), and jokes ("Factitious diarrhea: A case of watery deception", "HT06, tagging paper, taxonomy, Flickr, academic article, to read"). Thus, poetic phrases may be far more common and with different disciplinary variations than the relatively standardised versions found here. In addition, poetic metaphors can be shorter or even a single word (e.g., cuckoo), but it was impractical to check shorter phrases. As mentioned in the Methods, the manual checking stage is subjective and will have missed some poetic phrases. This may have influenced the conclusions from the results if poetic phrases from the natural sciences and engineering were substantially more likely to be overlooked. This might have happed if these subjects made relatively more frequent reference to areas of specialty or cultural knowledge unknown to the current article's author.

The results show that poetic phrases of multiple types are widely used across academic fields. This reveals, for the first time, that such phrases are relatively common in the social sciences and humanities, and confirm that some are common in medicine (Goodman, 2005, 2012; Gornitzki et al., 2015). Despite previous studies mainly focusing on biomedical research, however, the 149 stock phrases analysed here are far more common in social sciences, arts and humanities (Figure 1). The results also demonstrate that medium or long poetic phrases are rare in the natural sciences, formal sciences and engineering. Unless other types of poetic phrase (e.g., shorter, more heavily modified), do not follow this trend, it seems that the social sciences, arts and humanities are the natural home of poetic title phrases and that they are comparatively absent from the natural sciences, formal sciences and engineering.

This study has not analysed structural considerations behind poetic language use, such as whether colons are frequently used to separate nonliteral and literal segments of tiles, whether journals with stricter refereeing are less likely to accept them, or whether their use has increased over time. Information about these would give deeper insights into the phenomenon.

There are many possible causes of the apparent disciplinary differences in the prevalence of poetic titles. The relative abundance of poetic phrases in the social sciences and humanities journal titles may be influenced by journal guidelines, advice (e.g., Norman, 2012; Rossi & Brand, in press) or instructions (e.g., maximum title lengths). Three possible major causes are discussed below, starting with the one that seems to be the most important.

*A greater need to generate engaging titles to attract an audience*: In more linear science fields, papers may tend to be adding pieces to the jigsaw of the field, so people needing the information will find it by appropriate keyword searches (e.g., "Increasing the illumination slowly over several weeks protects against light damage in the eyes of the crustacean *Mysis* relicta"). In other areas, the knowledge may be optional rather than essential (e.g., "'What's love got to do with It?' The experience of love in person-product relationships", "Intellectuals and power, or, what's love got to do with it?"), and so engaging a browsing audience is more important. This is also the case for newspaper headlines or magazine titles, for example, which often use puns or other humour to attract attention (Alexander, 1986; Chovanec, 2005; Monsefi & Mahadi, 2016). The increased use of question marks in natural and life sciences over a decade ago has been attributed to an increasing need to market articles, however (Ball, 2009), but a greater proportion of social science articles might benefit from active marketing through titles. Related to this, in the social sciences, the diversity of human experience means that a theory cannot be comprehensively evaluated but can be tested in many different environments. In this situation, the author needs to make the case for the interestingness of the environment examined to attract an audience given that their contribution is unlikely to be individually vital to the theory. Incorporating an interesting quote in the title is another common strategy to achieve this goal (e.g., "Recommended for you" in: Hallinan & Striphas, 2016).

*A greater perceived freedom to experiment with article titles*: Historically-developed disciplinary norms, perhaps influenced by the factors above, may be considered to be transgressed if a title is not literal enough. For example, the inclusion of a pun may suggest that the article is less scientific and more like a magazine article. Closely related to this, new scholars may attempt to imitate current article title styles (Nagano, 2015), perhaps avoiding humour if they do not find evidence that this would be acceptable.

*A lesser need to succinctly summarize the contents of an article*: social sciences and humanities findings may be discursive or not easily summarised, whereas in the natural sciences and some areas of medicine, simple summarisations are possible (e.g., "Spectral Sensitivity of Single Photoreceptor Cells in the Eyes of the Ctenid Spider Cupiennius salei Keys", "Colony-stimulating factor-1 injections improve but do not cure skeletal sclerosis in osteopetrotic (op) mice").

# 6 Conclusion

Although the results are subjective to the cultural and subject knowledge of the author, the apparent relative scarcity of standard poetic phases in the natural sciences, formal sciences and engineering suggests that their use is against disciplinary norms in these areas. This does not prove that their use should be avoided because (a) the methods used here are not exhaustive, (b) it is possible that reviewers would be *more* receptive to the novelty of poetic or nonliteral titles in fields where they are scarce, and (c) no evidence is available about the perception of poetic expressions in any field. Nevertheless, academia is imitative and tribal through disciplinary cultures (Becher & Trowler, 2001; Hyland, 2012). Reviewers and readers may therefore judge disciplinary competence partly through adherence to writing style norms. Authors in relevant fields should therefore be particularly cautious when considering a common poetic phrase for their article titles, in case readers and reviewers form a negative judgement and their less keyword-rich title is less findable to other scholars. Of course, reviewers should be forgiving of authors that are not native English speakers using a poetic expression without realising that it is a cliché.

A likely cause of the disciplinary differences is a greater need to attract an audience in the social sciences, arts and humanities due to their inherent variety and non-hierarchical nature. This suggests that the construction of an engaging title is particularly important for authors in these areas, who should also consider alternative strategies, such as interesting quotes and questions (Ball, 2009; Cook & Plourde, 2016).

Finally, whilst the 149 poetic phrases analysed here occur in about 0.1% of Social Sciences articles, when added to more ad-hoc idiomatic or poetic phrases, it seems likely that poetic titles are a rhetorical device that reviewers and readers would often recognise as an accepted strategy in this area. Nevertheless, clichés may alienate part of the audience. In this context, novel poetic expressions might give the best of both worlds.

**DATA AVAILABILITY**
The data used for the graphs and the complete lists of n-grams, including those judged to be literal, is at: https://doi.org/10.6084/m9.figshare.12016278.

16